\documentclass[12pt]{iopart}

\usepackage{graphicx,xcolor}
\usepackage{cite}
\usepackage{marginnote}
\expandafter\let\csname equation*\endcsname\relax
\expandafter\let\csname endequation*\endcsname\relax
\usepackage{amsmath, amsthm, amssymb}
\usepackage{booktabs}
\usepackage{siunitx}
\UseRawInputEncoding

\newcommand{\ie}{{\emph{i.e.}}}
\usepackage{hyperref}
\hypersetup{colorlinks=true,linkcolor=black}
\begin{document}
\title{Resistive wall mode induced disruptions in an advanced tokamak}

	
	\author{Sui Wan}
	
	\address{State Key Laboratory of Advanced Electromagnetic Technology, \\International Joint Research Laboratory of Magnetic Confinement Fusion and Plasma Physics, School of Electrical and Electronic Engineering,
		\\	Huazhong University of Science and Technology, Wuhan, 430074,
		China}
	
	\author{Ping Zhu*}
	
	\address{State Key Laboratory of Advanced Electromagnetic Technology, \\International Joint Research Laboratory of Magnetic Confinement Fusion and Plasma Physics, School of Electrical and Electronic Engineering,
		\\	Huazhong University of Science and Technology, Wuhan, 430074,
		China;
		~\\
		Department of Nuclear Engineering and Engineering Physics, 
		\\University of Wisconsin-Madison, Madison,
		Wisconsin, 53706, United States of America}
	\ead{zhup@hust.edu.cn}
	\vspace{10pt}
	\begin{indented}
		\item[] \today
	\end{indented}
	\begin{abstract}
		Resistive wall mode is one of the leading causes for tokamak disruptions above the no-wall $\beta_N$ limit. This paper presents nonlinear three-dimensional resistive MHD simulations of an RWM-induced disruption in a CFETR baseline steady-state equilibrium using the NIMROD code. Linear calculations confirm the dominant presence of the $n=1$ RWM instability, whose growth rate is strongly sensitive to the wall response and becomes weakly dependent on plasma resistivity in the high-$S$ limit, along with a global external-kink-like structure. In the nonlinear phase, the RWM drives rapid flux surface stochastization and a thermal quench, followed by a current quench that is intensified by the post quench increase of Spitzer resistivity. The transient current spike before the current quench is shown to be the outcome of the conservation of poloidal flux and a rapid reduction of internal inductance. During the late current quench stage, closed flux surfaces partially reform from the core region to the edge, relaxing toward the force-free state. Toroidal mode coupling, parallel heat transport, plasma resistivity, and wall conductivity strongly modulate the disruption onset and the quench dynamics. Within the MHD model, these results provide a complete view on the RWM-driven disruption process in advanced tokamak configurations.
	\end{abstract}

	%
	\vspace{2pc}
	\noindent{\it Keywords}: nonlinear MHD, CFETR, resistive wall modes, NIMROD
	%
	
	\submitto{\PPCF}
	%
	%
	%
	\section{Introduction}\label{sec1}
	The development of advanced tokamak (AT) operating scenarios relies on plasmas with high poloidal beta ($\beta_p$) and a high fraction of non-inductive bootstrap current~\cite{takenaga07a,joffrin07a,greenfield04a,buttery19a,lucefor05a,tuccillo06a}. These high-performance regimes typically require operation at a normalized beta, $\beta_N$, that exceeds the no-wall stability limit calculated for ideal magnetohydrodynamic (MHD) modes~\cite{bondeson94a}. In addition, the reversed magnetic shear often present in the plasma core of AT scenarios can further destabilize low-$n$ modes~\cite{huysmans99a,takeji02a,turnbull02a,wan24a}. In this regime, the resistive wall mode (RWM)~\cite{finn95a,chu10a,garofalo07a} grows on the relatively slow resistive diffusion timescale of the wall ($\tau_w$), and if left uncontrolled, can lead to a major disruption, posing a severe threat to machine integrity. In machines operating above the ``no-wall" stability limit, the RWM is the primary instability that must be controlled to avoid sudden termination~\cite{garofalo00a}. Global MHD instabilities such as RWMs can be the root cause of the most rapid disruptions with the least amount of pre-disruption warning time~\cite{sabbagh23a}. In an NSTX database study of unstable RWMs, the onset of the mode led to a disruption event in 100\% of cases~\cite{sabbagh23a}.

	For CFETR~\cite{han20a, zhu22a, han23a, wan26a} specifically, ideal MHD calculations indicate that several CFETR steady-state equilibria lie above the no-wall $\beta_N$ limit, becoming unstable to an $n=1$ external kink if no perfectly conducting wall is assumed. Consequently, the RWM emerges as a key active stability concern for CFETR. Linear studies have evaluated the RWM growth rates and thresholds in these scenarios, including the beneficial effects of kinetic damping and plasma flow on RWM stability~\cite{li22a, li22b}. Despite this progress, nonlinear investigations of RWM dynamics remain comparatively limited. Previous nonlinear calculations have primarily served as physics-demonstration studies under idealized assumptions (e.g., simplified geometry and boundary representations)~\cite{sato03a,finn04a,sato06a,sato06b}, whereas fully nonlinear, device-realistic simulations that resolve the complete disruption process as the consequence of RWMs remain rare. Most existing nonlinear simulations related to CFETR have focused on specific mitigation techniques or alternate disruption triggers~\cite{tang22a,zeng25a, zeng26a}, rather than examining the physics of an RWM-induced collapse in an unmitigated steady-state plasma. The characteristics of the RWM-led disruption process as well as its transient and terminal influence has been less well known.
	
	In this paper, we address this gap by performing fully nonlinear resistive MHD simulations of an RWM-induced disruption in CFETR baseline steady-state scenario using the NIMROD code~\cite{sovinec04a}. The disruption is driven by a global unraveling of flux surfaces, leading to a rapid thermal collapse. Furthermore, during the ensuing current quench phase, a spontaneous magnetic flux healing phenomenon is observed. We interpret this counter-intuitive partial recovery of magnetic surfaces as a manifestation of Taylor relaxation~\cite{taylor86a} toward a lower-energy, force-free helical state.
	
	The remainder of this paper is organized as follows. Section \ref{sec:nimrod_model} describes the single-fluid resistive MHD model and the CFETR equilibrium configuration adopted for this study. Section \ref{sec3} presents the numerical results, on the linear verification of the RWM, the nonlinear dynamics of the TQ, and the phenomenon  of post-disruption flux healing, alongside sensitivity scans of key parameters. Finally, Section \ref{sec4} summarizes the findings and discusses their implications for CFETR operations.
	
	\section{Numerical model and equilibrium}
	\label{sec:nimrod_model}
	
	We adopt the single-fluid resistive MHD model implemented in the NIMROD code~\cite{sovinec04a} that includes viscosity and anisotropic thermal conduction. The governing equations are:
	
	\begin{gather}
		\frac{\partial N}{\partial t} + \nabla \cdot (N\boldsymbol{u})
		= \nabla \cdot (D \nabla N),
		\label{eq:nimrod_continuity}
		\\
		m_i N\left(\frac{\partial \boldsymbol{u}}{\partial t} + \boldsymbol{u}\cdot\nabla \boldsymbol{u}\right)
		= \boldsymbol{J}\times\boldsymbol{B} - \nabla p - \nabla\cdot\boldsymbol{\Pi},
		\label{eq:nimrod_momentum}
		\\
		\frac{N}{\Gamma-1}\left(\frac{\partial T}{\partial t} + \boldsymbol{u}\cdot\nabla T\right)
		= -p\,\nabla\cdot\boldsymbol{u}-\nabla \cdot \boldsymbol{q} + Q,
		\label{eq:nimrod_temperature}
		\\
		\frac{\partial \boldsymbol{B}}{\partial t}
		= -\nabla\times\left(\eta\boldsymbol{J} - \boldsymbol{u}\times\boldsymbol{B}\right)
		,
		\label{eq:nimrod_induction}
		\\
		\mu_0 \boldsymbol{J} = \nabla\times\boldsymbol{B},
		\label{eq:nimrod_ampere}
		\\
		\boldsymbol{q} = -N\left[ \kappa_\parallel \hat{\boldsymbol{b}}\hat{\boldsymbol{b}}\cdot\nabla T + \kappa_\perp\left(\boldsymbol{I}-\hat{\boldsymbol{b}}\hat{\boldsymbol{b}}\right)\cdot\nabla T\right] ,
		\label{eq:nimrod_conduction}
	\end{gather}
	where $N$ is the plasma number density, $\boldsymbol{u}$ is the center-of-mass flow velocity, $T$ is the plasma temperature ($T=T_i=T_e$), and $p$ is the scalar pressure. The magnetic field and current density are denoted as $\boldsymbol{B}$ and $\boldsymbol{J}$, respectively. The constant definitions include the ion mass $m_i$, the vacuum permeability $\mu_0$, and the adiabatic index $\Gamma=5/3$. The term $D$ represents an artificial number density diffusivity, and $Q$ encompasses energy source terms. The viscous stress tensor is denoted as $\boldsymbol{\Pi}$. The plasma resistivity follows the Spitzer model $\eta=\eta_0\left(T_0/T \right)^{3/2}$~\cite{spitzer53a,kuritsyn06a}, where $T_0$ is the reference temperature at the magnetic axis. The heat flux $\boldsymbol{q}$ in Eq.~\eqref{eq:nimrod_conduction} accounts for anisotropic thermal transport, where $\hat{\boldsymbol{b}}=\boldsymbol{B}/|\boldsymbol{B}|$ is the unit vector along the magnetic field line. The coefficients $\kappa_\parallel$ and $\kappa_\perp$ are the parallel and perpendicular thermal conductivities relative to the magnetic field, respectively. Given that typically $\kappa_\parallel \gg \kappa_\perp$, this anisotropic model is critical for capturing the rapid parallel heat loss once magnetic flux surfaces become stochastic during the disruption process.
	
\begin{figure}[h]
		\centering
		\includegraphics[width=0.7\linewidth]{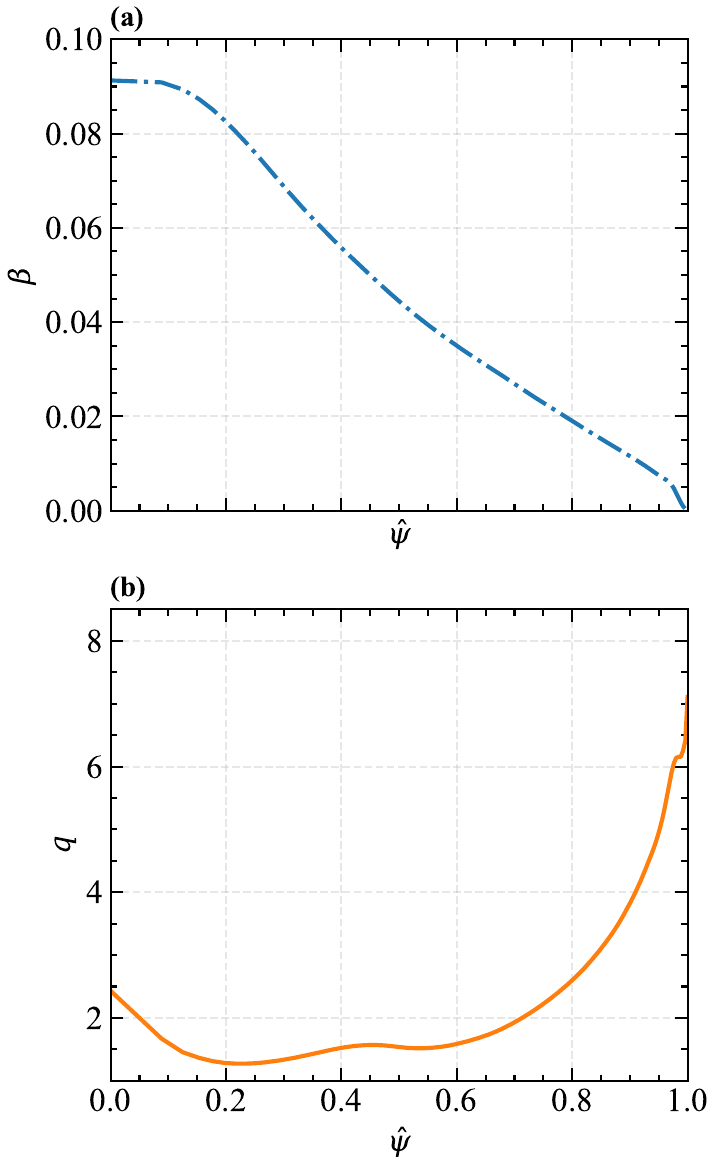}
		\caption{(a) Equilibrium $\beta$ and (b) safety factor $q$ profiles as functions of the normalized poloidal flux function $\hat{\psi}$ for \texttt{CFETR-1}.}
		\label{fig:eq}
	\end{figure}

	In this work, the equilibrium used for the nonlinear simulations is based on the CFETR baseline steady-state scenario~\cite{zhuang19a,chen21a,chen21b,zhou22a}. The equilibrium, referred to as \texttt{CFETR-1}, has a major radius $R_0=7.188$~m, a minor radius $a=2.231$~m, a toroidal magnetic field $B_0=6.541$~T, and a total plasma current $I_p=13.3$~MA. This scenario features a weakly negative magnetic shear in the core region. The equilibrium profiles are shown in Fig.~\ref{fig:eq}, and the key equilibrium parameters are summarized in Table~\ref{tab:equilibrium_params}.
	
	\begin{table}[htbp]
		\centering
		\caption{Key parameters of the equilibrium}
		\label{tab:equilibrium_params}
		\begin{tabular}{llll}
			\toprule
			Parameter & Symbol & Value & Unit \\
			\midrule
			Minor radius & $a$ & 2.231 & m \\
			Major radius & $R_0$ & 7.188 & m \\
			Plasma current & $I_p$ & $13.33$ & MA \\
			Toroidal magnetic field & $B_0$ & 6.541 & T \\
			Central safety factor & $q_0$ & 2.414 & Dimensionless \\
			Edge safety factor & $q_{99}$ & 7.040 & Dimensionless \\
			Temperature at magnetic axis & $T_{\text{e,core}}$ & $17.48$ & keV \\
			Number density at magnetic axis & $n_{\text{e,core}}$ & $2.400 \times 10^{20}$ & m$^{-3}$ \\
			\bottomrule
		\end{tabular}
	\end{table}
	
To represent the boundary conditions relevant to RWM physics, a double-boundary domain with two vacuum regions is employed. The inner boundary representing the resistive wall separates the inner pseudo-vacuum region from the outer region, which is surrounded by the outer boundary with an ideal wall condition. When the outer ideal wall is placed sufficiently far from the plasma, its influence on the unstable mode is negligible, similar to the previous NIMROD simulations on vertical displacement events~\cite{sovinec19a,bunkers20a,li24a}. The response of the resistive wall on the inner boundary is described using the thin-wall approximation in terms of the normal magnetic field to the wall
\begin{equation}
	\frac{\partial\left(\boldsymbol{B}\cdot\hat{\boldsymbol{n}}\right)}{\partial t}
	=-\hat{\boldsymbol{n}}\cdot\nabla\times\left[v_w\,\hat{\boldsymbol{n}}\times\Delta\boldsymbol{B}\right],
	\label{eq:nimrod_thinwall_bn}
\end{equation}
where $\hat{\boldsymbol{n}}$ is the outward unit normal vector, and $\Delta\boldsymbol{B}$ denotes the jump in the tangential component of magnetic field across the wall. The wall properties are parameterized using the diffusion velocity $v_w=\eta_w/(\mu_0\delta_w)$, with $\eta_w$ and $\delta_w$ being the wall resistivity and thickness, respectively.

\begin{figure}[h]
		\centering
		\includegraphics[width=1.2\linewidth]{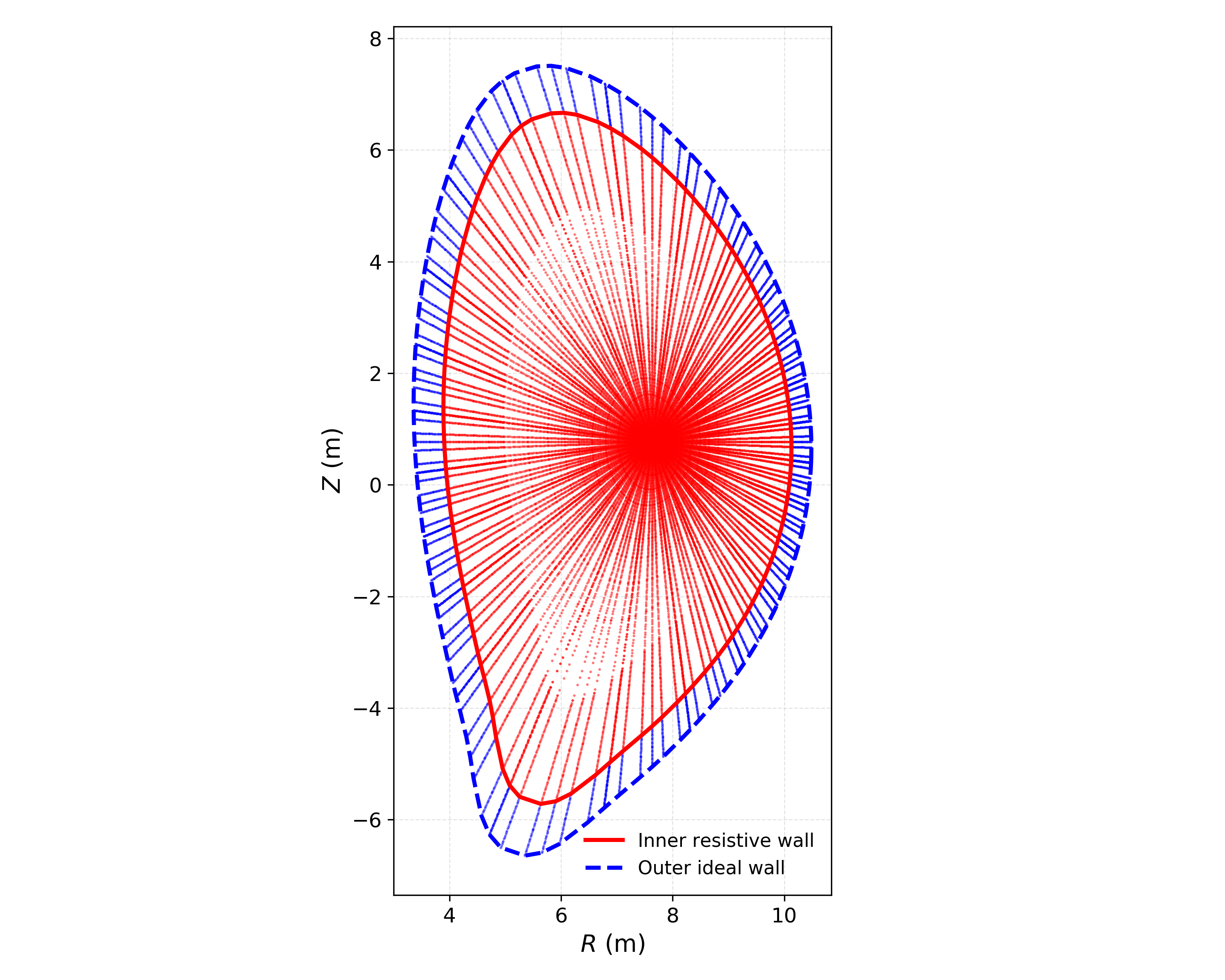}
		\caption{Schematic of the double-wall poloidal domain and grid points used in NIMROD simulations. The red region denotes the plasma and the inner vacuum enclosed by the inner resistive wall, while the blue region represents the outer vacuum region between the two walls.}
		\label{fig:nimrod-double-wall-grid}
	\end{figure}

For the \texttt{CFETR-1} simulations in this work, the inner resistive wall is conformal to the plasma boundary and located at $r_{w,\mathrm{in}}/a=1.4$, where $a$ is the plasma minor radius. The outer ideal wall is placed at $r_{w,\mathrm{out}}/a=1.6$ and serves as the outmost boundary of the entire computational domain (Fig.~\ref{fig:nimrod-double-wall-grid}).
	
	\section{Numerical results}\label{sec3}
	\subsection{Linear RWM growth and structure}\label{subsec:linear_verification}
	
\begin{figure}[h]
		\centering
		\includegraphics[width=0.8\linewidth]{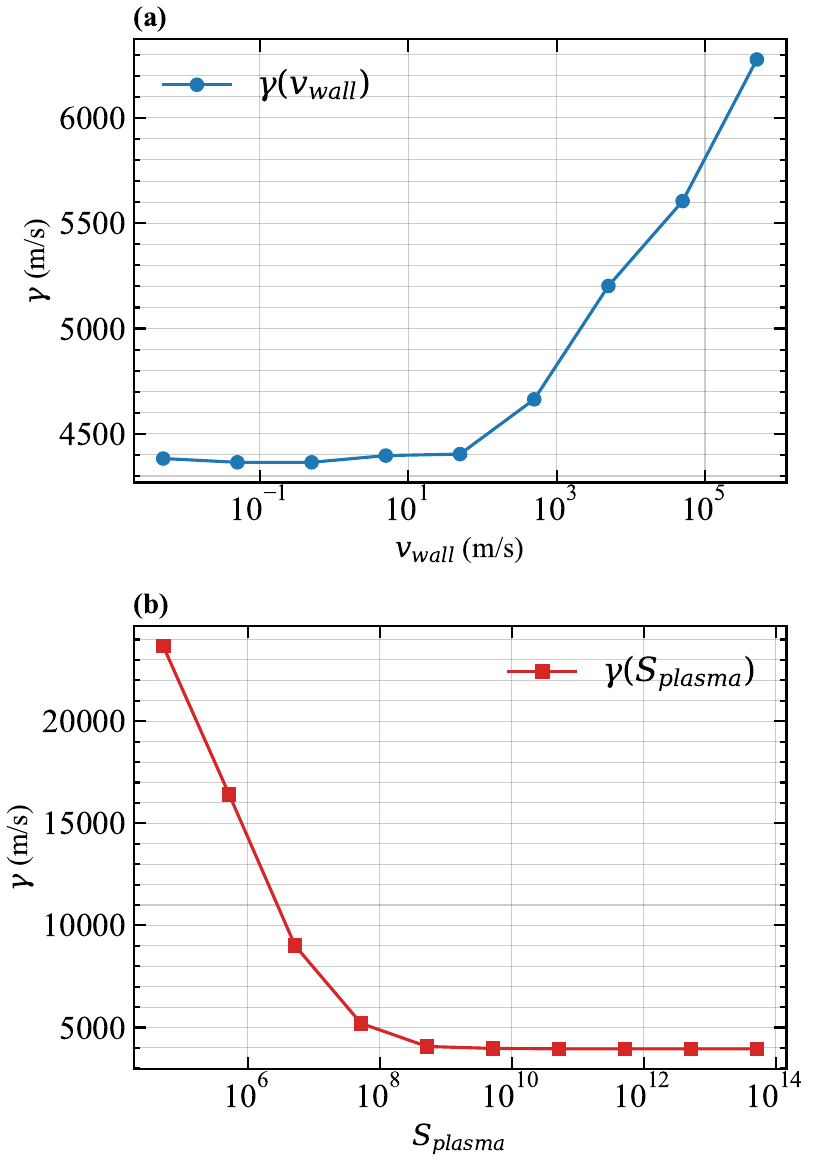}
		\caption{Linear growth rates of the $n=1$ mode as functions of (a) the wall diffusion velocity $v_w$ and (b) the plasma Lundquist number $S$.}
		\label{fig:eq1-linear-gamma}
	\end{figure}

\begin{figure}[h]
		\centering
		\includegraphics[width=0.5\linewidth]{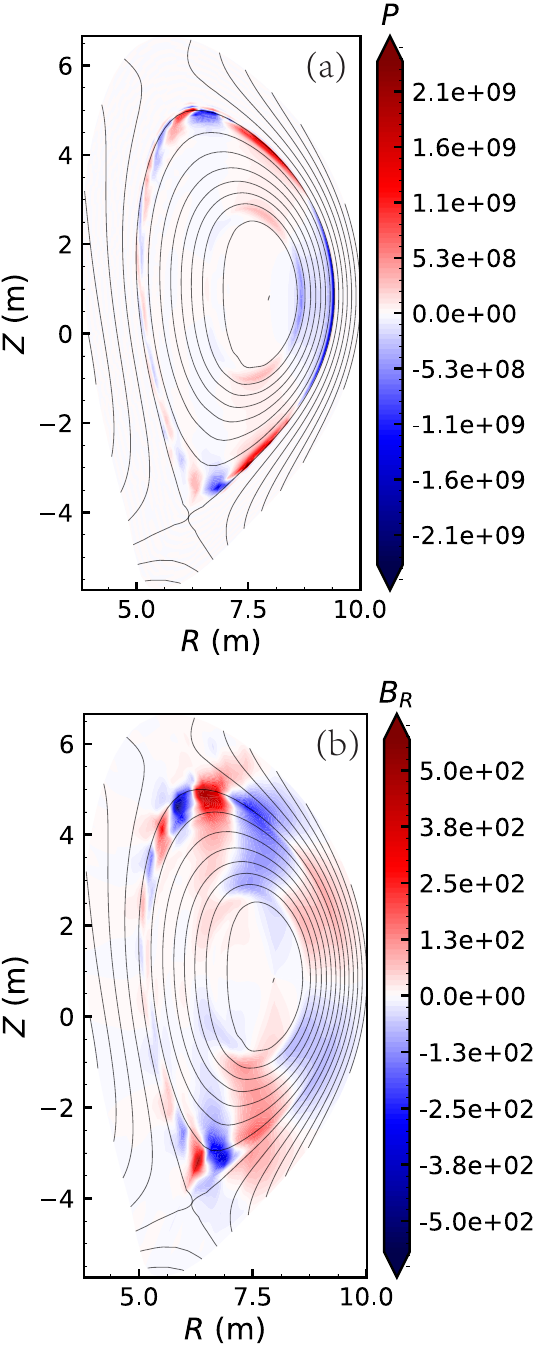}
		\caption{Contours of the (a) perturbed pressure and (b) perturbed radial component of magnetic field from the linear $n=1$ mode (color) along with the poloidal magnetic flux in the poloidal plane.}
		\label{fig:eq1-linear-mode-structure}
	\end{figure}

	A series of linear simulations were performed to examine the dependence of the growth rate $\gamma$ on the wall response and plasma resistivity. As shown in Fig.~\ref{fig:eq1-linear-gamma}(a), the growth rate increases monotonically with the characteristic wall diffusion velocity, demonstrating a strong dependence on the finite conductivity of the wall. As depicted in Fig.~\ref{fig:eq1-linear-gamma}(b), the growth rate $\gamma$ approaches an asymptotic value for $S \gtrsim 10^8$, indicating the ideal MHD nature of the instability. The mode structure further supports an external kink-type character. The $n=1$ perturbed pressure in Fig.~\ref{fig:eq1-linear-mode-structure}(a) is global and peaks near the plasma edge. The corresponding perturbed radial component of the magnetic field in Fig.~\ref{fig:eq1-linear-mode-structure}(b) remains finite at the plasma boundary and extends into the vacuum region, indicating strong electromagnetic coupling to the surrounding conducting wall. The linear growth dependence and mode structure agree with those of an RWM.
	
	\subsection{Nonlinear evolution and disruption}\label{subsec:nonlinear_cfetr1}
	 To assess the impact of toroidal mode coupling on the disruption dynamics, two nonlinear simulations were performed with different toroidal spectral truncations: one retaining only the dominant low-$n$ components ($n=0$--$1$) and another including higher-order harmonics ($n=0$--$5$). Unless otherwise specified, the following nonlinear simulations use $v_{\mathrm{wall}}=5\times10^{3}~\mathrm{m\,s^{-1}}$ and $S = 4.7\times10^{9}$.
	
\begin{figure}[h]
		\centering
		\includegraphics[width=1\linewidth]{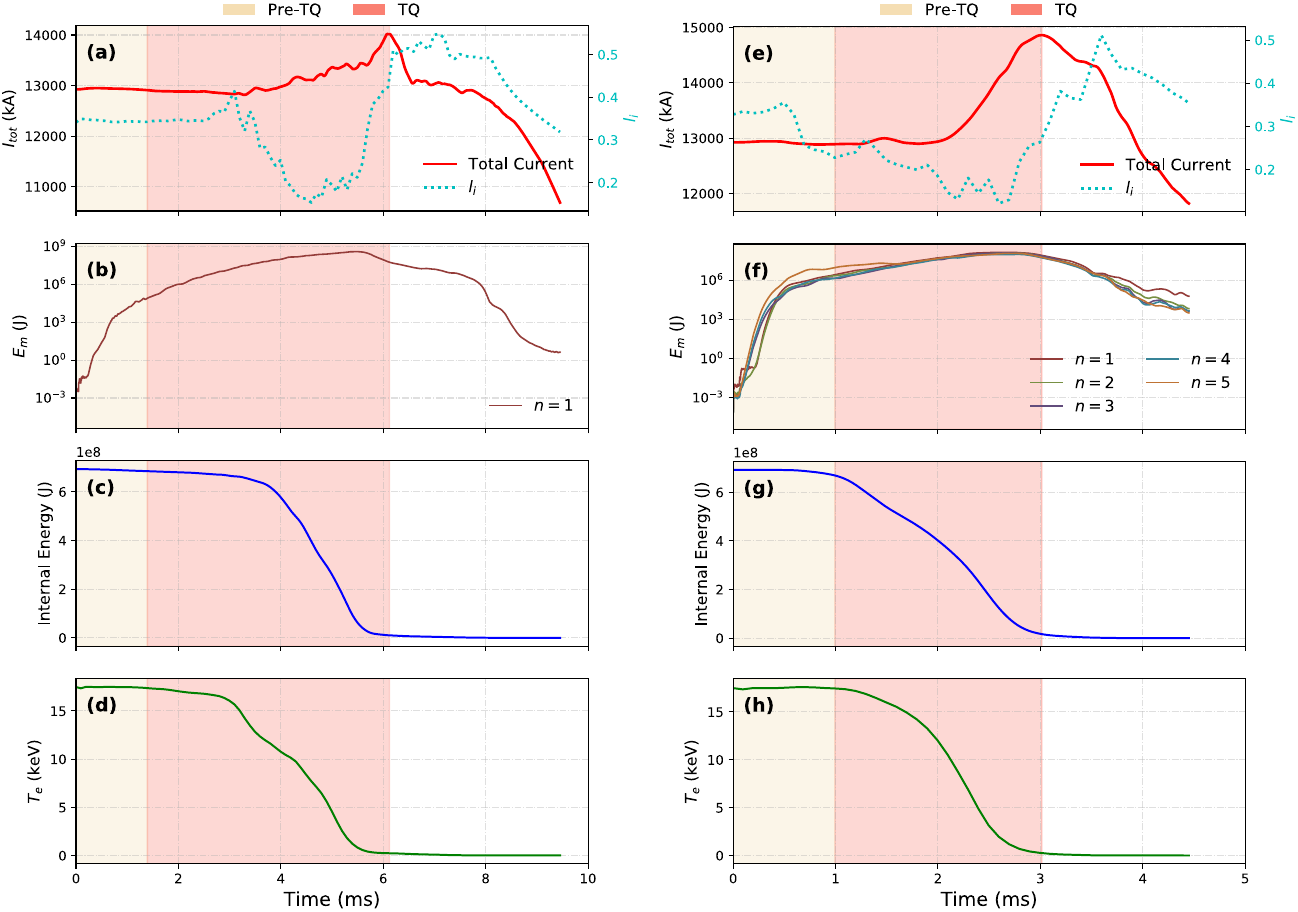}
		\caption{(a, e) Total plasma current $I_p$ (solid red line) and internal inductance $l_i$ (dotted blue line), (b, f) magnetic energy of different toroidal modes ($n \ge 1$), (c, g) total plasma internal energy, and (d, h) electron temperature at the magnetic axis as functions of time. The left column (a-d) corresponds to the simulation with toroidal mode components $n=0-1$, while the right column (e-h) corresponds to the case including $n=0-5$. The yellow shaded time interval indicates the pre-thermal quench (Pre-TQ) phase, and the pink shaded time interval indicates the thermal quench (TQ) phase.}
		\label{fig:eq1-nonlinear-dis}
	\end{figure}

	 Fig.~\ref{fig:eq1-nonlinear-dis} presents the time evolution of RWM and the subsequent disruption in three phases: pre-thermal quench (Pre-TQ), thermal quench (TQ), and current quench (CQ). During the Pre-TQ phase, the magnetic perturbation grows exponentially approximately (Fig.~\ref{fig:eq1-nonlinear-dis}(b,f)), even though the core confinement remains largely preserved, as indicated by the nearly constant electron temperature $T_e$ at the magnetic axis and total internal energy (Fig.~\ref{fig:eq1-nonlinear-dis}(c,d,g,h)). 
	 
	 The onset of the TQ marks a macroscopic collapse. Once the magnetic perturbations reach a sufficiently large amplitude, the magnetic topology becomes globally stochastic, opening efficient parallel transport channels and triggering a rapid loss of confinement. Consequently, the core temperature collapses to below 10\% of its initial value on a millisecond timescale. Meanwhile, the perturbed magnetic level reaches its maximum. A characteristic current spike is observed near the end of the TQ (Fig.~\ref{fig:eq1-nonlinear-dis}(a,e)), accompanied by a pronounced reduction in internal inductance. This behavior is consistent with the approximate conservation of poloidal magnetic flux on the fast TQ timescale, where the resistive diffusion remains limited. Following the TQ, the plasma enters the CQ phase. The dramatic reduction in $T_e$ leads to a sharp rise in plasma resistivity according to the Spitzer scaling $\eta \propto T_e^{-3/2}$, and the resulting strong resistive dissipation drives the irreversible decay of the plasma current and magnetic energy.
	 
\begin{figure}[h]
		\centering
		\includegraphics[width=0.7\linewidth]{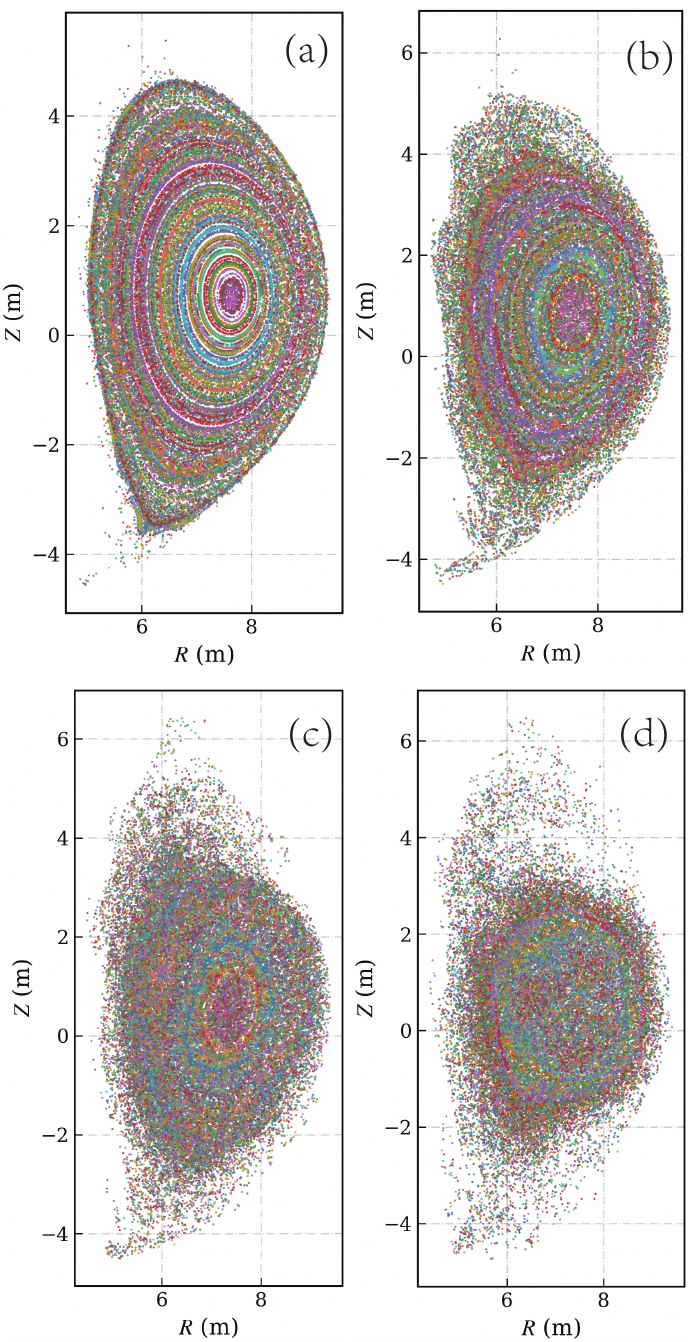}
		\caption{Poincar\'e plots in the poloidal plane at various times: (a) 1.5~ms, (b) 3.1~ms, (c) 3.8~ms, and (d) 4.5~ms.}
		\label{fig:eq1-poincare}
	\end{figure}

\begin{figure}[h]
		\centering
		\includegraphics[width=0.8\linewidth]{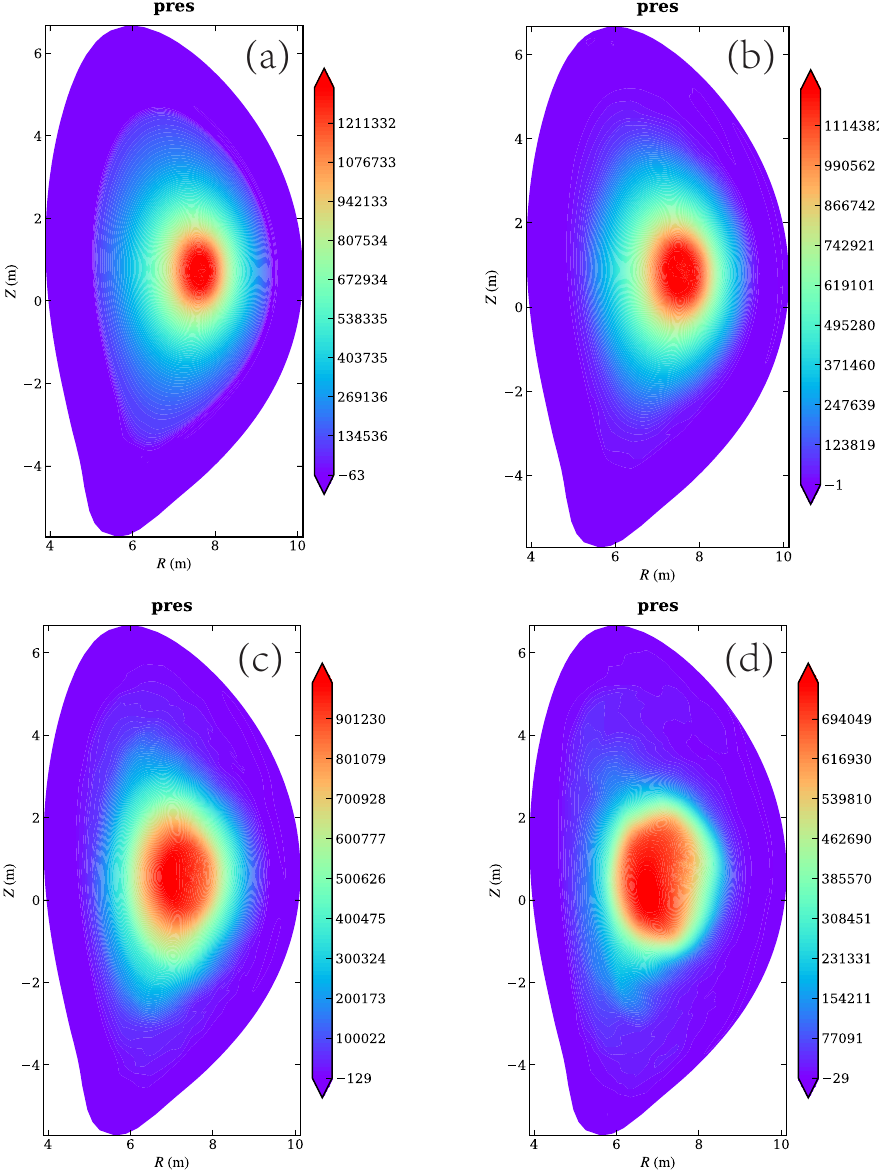}
		\caption{Total pressure contours (Pa) in the poloidal plane at various times: (a) 1.5~ms, (b) 3.1~ms, (c) 3.8~ms, and (d) 4.5~ms.}
		\label{fig:eq1-pres-contour}
	\end{figure}

	Unlike the classical tearing-mode picture where localized islands grow and overlap, the RWM in \texttt{CFETR-1} exhibits a more global and collective destruction of flux surfaces. The flux surfaces are initially intact (Fig.~\ref{fig:eq1-poincare}(a)), but as the mode grows, they undergo strong global deformation, with pronounced twisting developing from the edge and propagating inward (Fig.~\ref{fig:eq1-poincare}(b,c)). This behavior is consistent with an external-kink-like structure. The disruption is therefore initially driven by a global unraveling of flux surfaces rather than by localized reconnection at individual rational surfaces. By the peak of the TQ (Fig.~\ref{fig:eq1-poincare}(d)), the Poincar\'e plot becomes broadly ergodized, providing the transport pathway for the rapid thermal collapse, which is consistent with the evolution of the total pressure contours (Fig.~\ref{fig:eq1-pres-contour}).
	\subsection{Post-disruption relaxation and magnetic flux healing}
	\label{subsec:flux_healing}
	
\begin{figure}[h]
		\centering
		\includegraphics[width=0.9\linewidth]{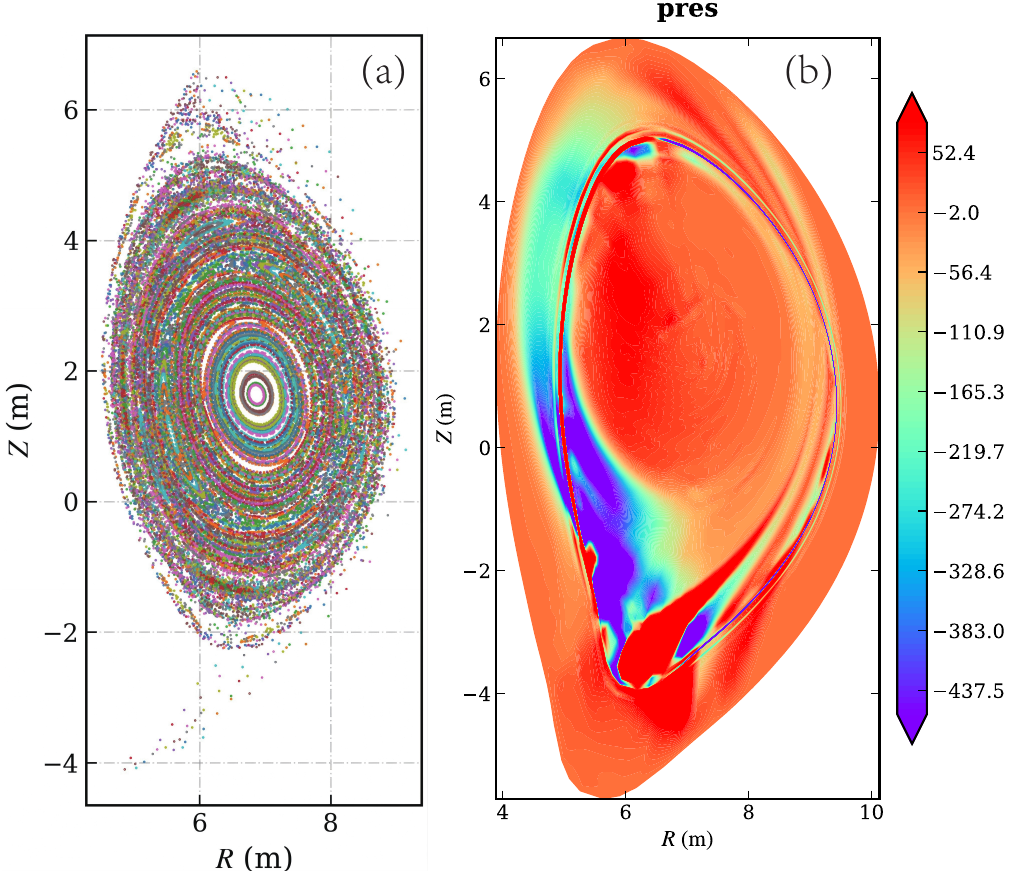}
		\caption{(a) Poincar\'e plots and (b) total pressure contours (Pa) in the poloidal plane at $t=8.1$ ms indicating the magnetic flux healing.}
		\label{fig:eq1-healing}
	\end{figure}

	A remarkable phenomenon observed in the deep current quench phase starting approximately from $t = 8.1$ ms in the reference simulation is the spontaneous ``healing'' of magnetic flux surfaces. As illustrated in Fig.~\ref{fig:eq1-healing}, magnetic field lines that are globally stochastic near the peak of the thermal quench progressively reorganize into closed flux surfaces, starting from the vicinity of the magnetic axis and expanding outward.
	
\begin{figure}[h]
		\centering
		\includegraphics[width=0.9\linewidth]{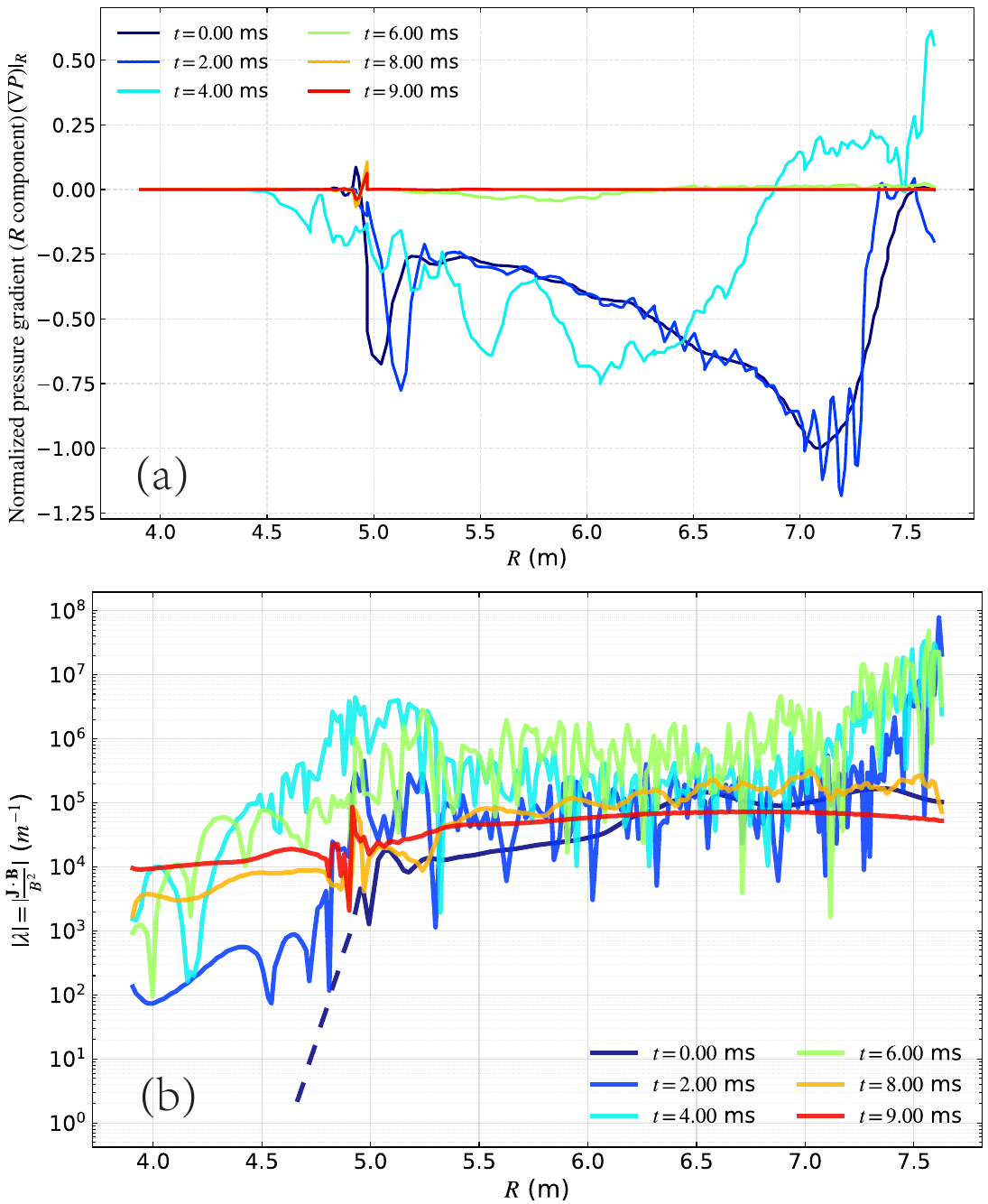}
		\caption{Radial profiles of (a) the normalized $R$-component of the pressure gradient and (b) the parallel current density parameter $\left| (\boldsymbol{J} \cdot \boldsymbol{B})/B^2 \right|$ as functions of the major radius $R$ in the midplane at various times.}
		\label{fig:eq1-gradp}
	\end{figure}

	This re-emergence of ordered topology is consistent with relaxation toward a near force-free equilibrium following the violent disruption. The underlying trend is revealed by the evolution of the radial profiles of the pressure gradient and the parallel current density, as shown in Fig.~\ref{fig:eq1-gradp}. During the disruptive phase, the pressure gradient $(\nabla p)|_R$ fluctuates strongly as stochastic transport rapidly redistributes pressure. In the subsequent relaxation stage, the pressure gradient effectively vanishes across the major radius (Fig.~\ref{fig:eq1-gradp}(a)). Assuming the quasi-static macroscopic MHD equilibrium condition $\boldsymbol{J} \times \boldsymbol{B} = \nabla p$, a vanishing $\nabla p$ implies a force-free configuration, \ie $\boldsymbol{J}\parallel\boldsymbol{B}$.
	
	This transition is further supported by the profile of the field-aligned current density, $|\boldsymbol{J} \cdot \boldsymbol{B} / B^2|$, shown in Fig.~\ref{fig:eq1-gradp}(b). Coincident with the vanishing pressure gradient, this quantity evolves toward a more flattened radial profile with a magnitude comparable to the initial equilibrium. Such flattening is consistent with a Taylor-like relaxation~\cite{taylor86a}, indicating that the plasma has reorganized into a minimum energy state subject to the conservation of global magnetic helicity, rather than dissipating currents uniformly to zero.
	
	The observation of magnetic flux healing is potentially relevant to post-disruption confinement. In particular, the partial re-formation of closed flux surfaces may provide a pathway for re-establishing confinement of runaway electrons (REs) generated during the quench, suggesting that simple stochastic loss models may overestimate RE decontamination in reactor-scale plasmas.
	
	\subsection{Impact of physical and numerical parameters on disruption}\label{subsec:sensitivity_analysis}
	
	While the qualitative sequence of the disruption remains robust across simulations, the specific timescales and severity of the event are sensitive to the toroidal spectral truncation and key transport/dissipation parameters. In this section, we systematically examine the impact of the toroidal mode truncation number, plasma resistivity, parallel thermal conductivity, and wall resistivity on the disruption evolution.
	
\begin{figure}[h]
		\centering
		\includegraphics[width=0.7\linewidth]{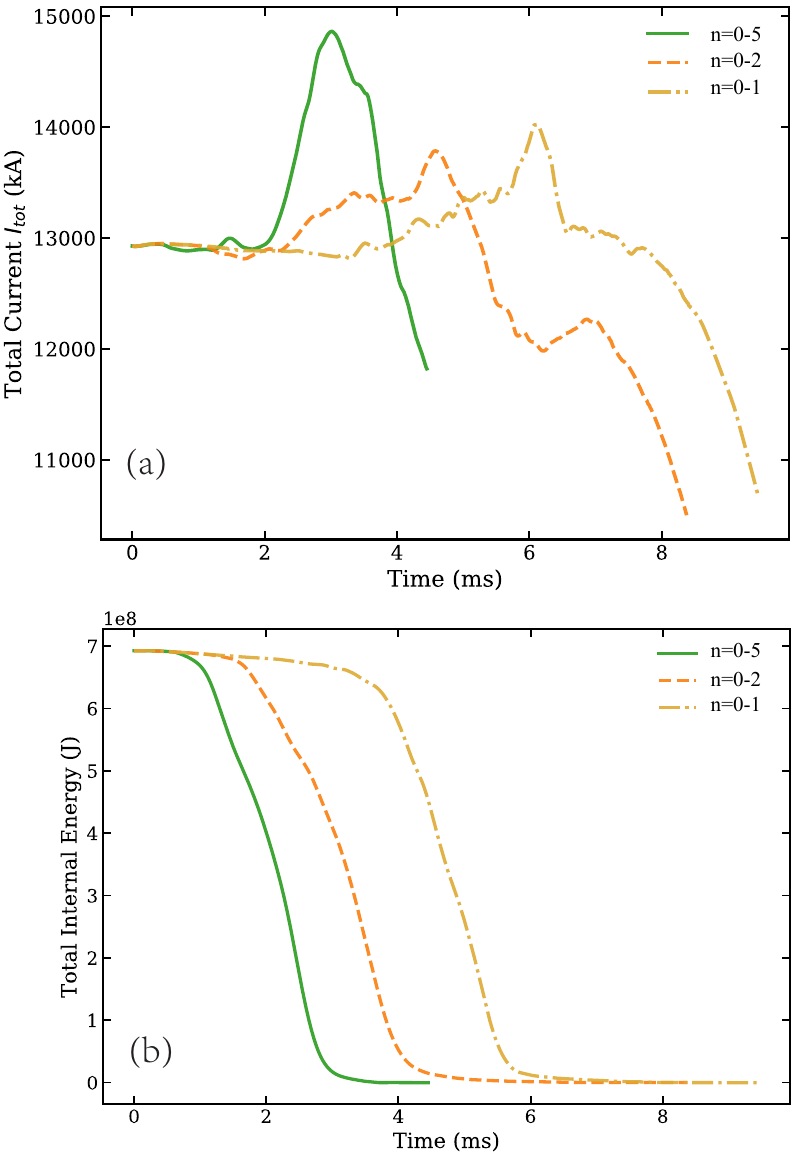}
		\caption{Time evolution of (a) the total plasma current and (b) the total internal energy for various toroidal mode truncations. The plasma Lundquist number $S=4.7\times 10^{9}$, the parallel to perpendicular heat conductivity ratio $\kappa_\parallel/\kappa_\perp=1\times 10^{8}$, and the wall penetration speed $v_{wall}=5\times 10^{3}$ m/s.}
		\label{fig:eq1-different-lphi}
	\end{figure}

	First, the nonlinear interaction between different toroidal harmonics plays a crucial role in the stochastization of the magnetic field. As shown in Figure~\ref{fig:eq1-different-lphi}, it is observed that as the number of included high-order harmonics increases, the onset of the TQ and CQ phases shifts significantly earlier. This acceleration can be attributed to the  energy cascade from the dominant low-$n$ modes to higher-$n$ harmonics via nonlinear coupling. These high-$n$ modes are more effective at breaking magnetic surfaces and creating ergodic regions, thereby accelerating the global degradation of confinement.
	
\begin{figure}[h]
		\centering
		\includegraphics[width=0.7\linewidth]{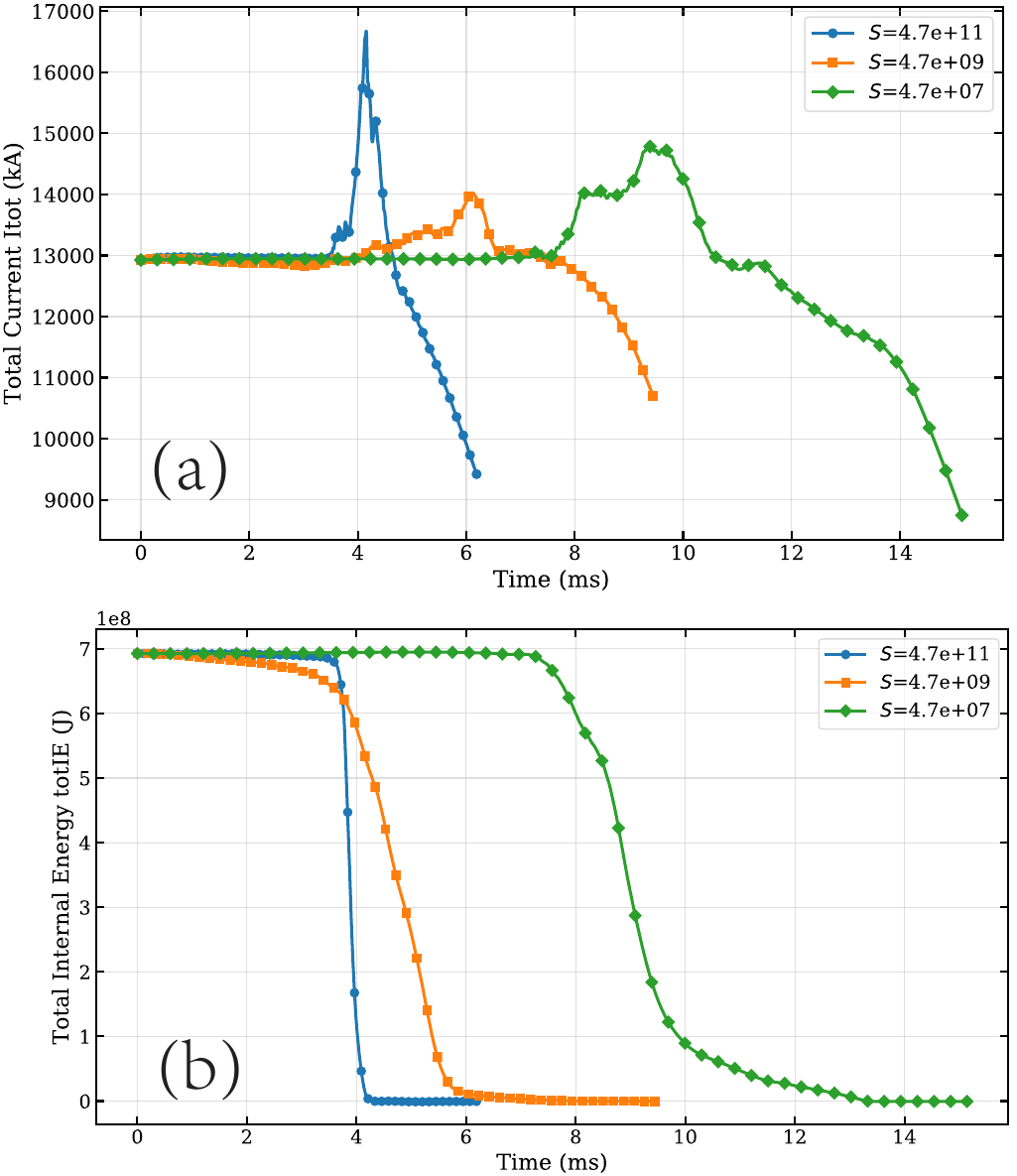}
		\caption{Time evolution of (a) the total plasma current and (b) the total internal energy for various plasma Lundquist numbers $S$. The included toroidal components are $n=0, 1$, the parallel to perpendicular heat conductivity ratio $\kappa_\parallel/\kappa_\perp=1\times 10^{8}$, and the wall penetration speed $v_{wall}=5\times 10^{3}$ m/s.}
		\label{fig:eq1-different-elecd}
	\end{figure}

	Second, the influence of plasma resistivity is evaluated by scanning the Lundquist number $S$. As shown in Fig.~\ref{fig:eq1-different-elecd}, the disruption process is notably delayed as the plasma resistivity increases (i.e., lower $S$). In the regime of interest, the RWM is an ideal-MHD driven instability. Enhanced plasma resistivity provides a background dissipative mechanism that acts against the ideal driving force, effectively slowing down the nonlinear growth of the mode. Consequently, simulations with higher resistivity exhibit a more protracted pre-TQ phase.
	
\begin{figure}[h]
		\centering
		\includegraphics[width=0.7\linewidth]{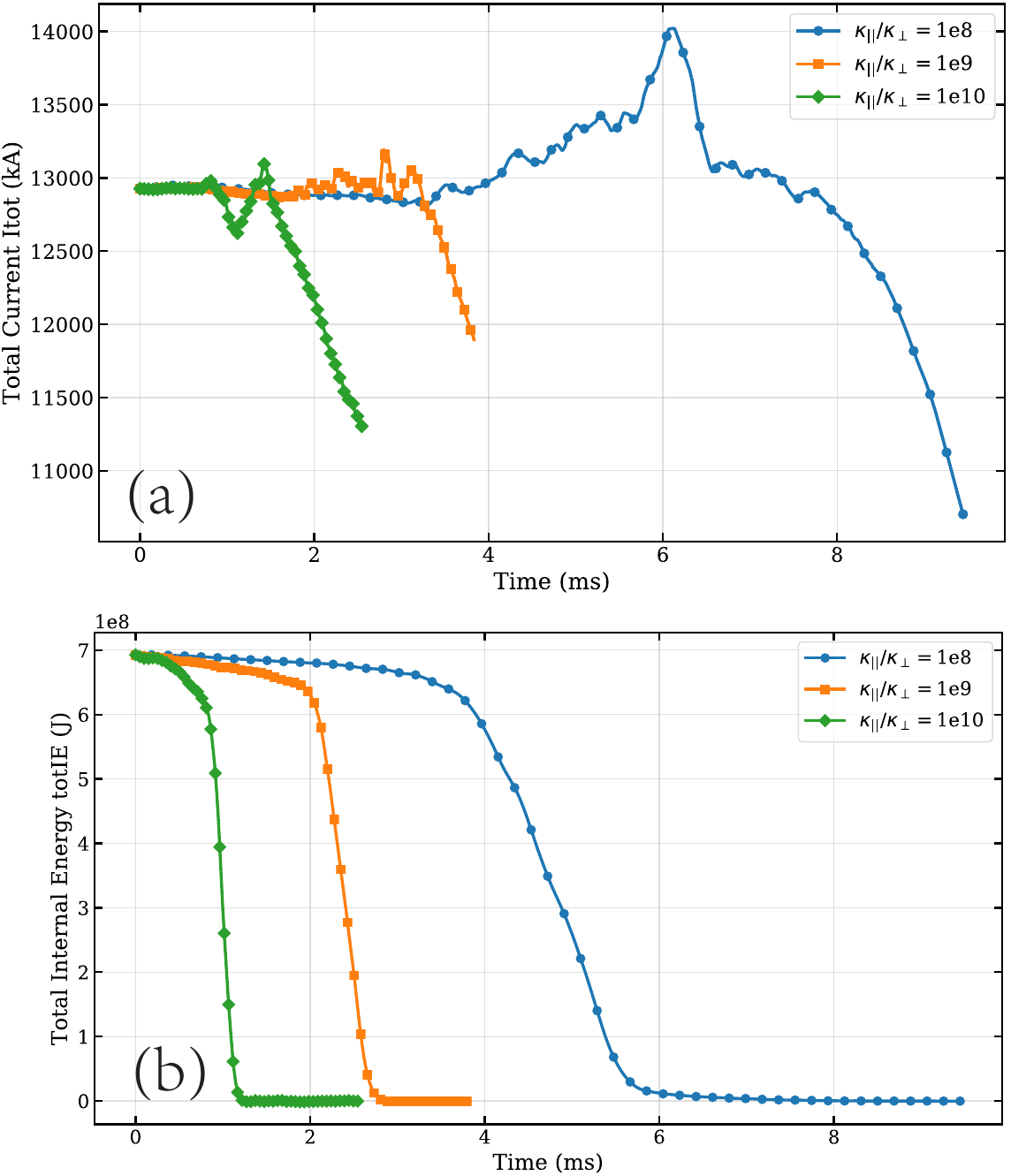}
		\caption{Time evolution of (a) the total plasma current and (b) the total internal energy for various parallel heat conductivity ratio $\kappa_\parallel/\kappa_\perp$. The included toroidal components are $n=0, 1$, the plasma Lundquist number $S=4.7\times 10^{9}$, and the wall penetration speed $v_{wall}=5\times 10^{3}$ m/s.}
		\label{fig:eq1-different-kpll}
	\end{figure}

	Third, the timescale of the thermal quench is directly governed by the parallel thermal transport physics. As shown in Figure~\ref{fig:eq1-different-kpll}, a higher $\kappa_\parallel/\kappa_\perp$ results in a more precipitous drop in internal energy. This is physically consistent with the Rechester-Rosenbluth transport model~\cite{rechester78a}; the effective radial heat transport increases with $\kappa_\parallel$ once the field becomes stochastic. Therefore, a larger $\kappa_\parallel$ facilitates more rapid heat exhaust along the open field lines to the wall, leading to a faster and more violent thermal quench.
	
\begin{figure}[h]
		\centering
		\includegraphics[width=0.7\linewidth]{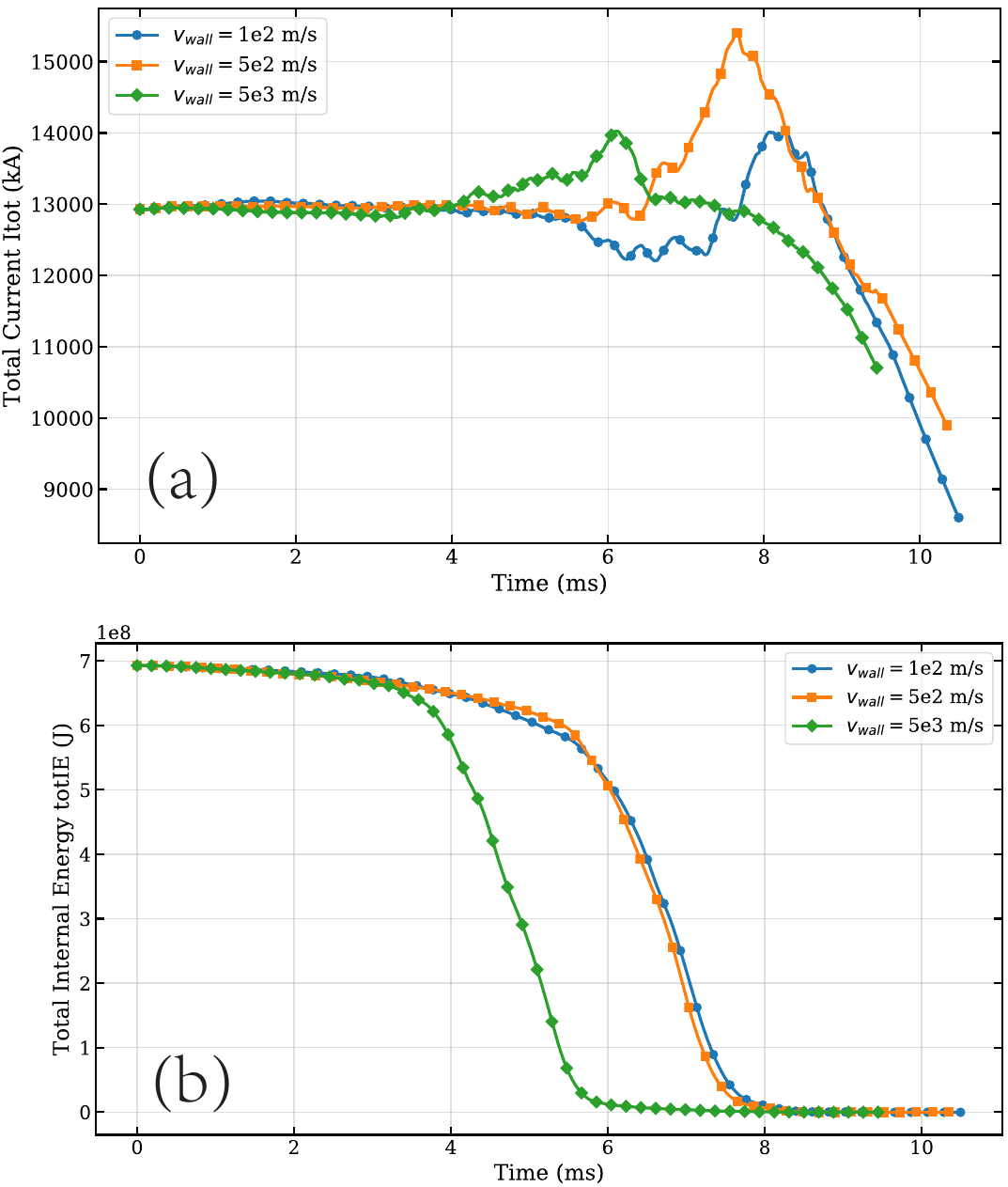}
		\caption{Time evolution of (a) the total plasma current and (b) the total internal energy for various wall penetration speed $v_{\mathrm{wall}}$. The included toroidal components are $n=0, 1$, the plasma Lundquist number $S=4.7\times 10^{9}$, and the parallel to perpendicular heat conductivity ratio $\kappa_\parallel/\kappa_\perp=1\times 10^{8}$.}
		\label{fig:eq1-different-vwall}
	\end{figure}

	Finally, the resistive wall penetration time is the fundamental scale governing the linear growth of the RWM. Figure~\ref{fig:eq1-different-vwall} demonstrates the dependence of the disruption timing on the characteristic wall diffusion velocity $v_w$. As the wall becomes more conductive (lower $v_w$, closer to the ideal wall limit), the eddy currents induced in the wall persist longer, effectively shielding the external magnetic perturbations. This shielding effect significantly reduces the mode growth rate, thereby delaying the onset of the nonlinear disruption. Conversely, a highly resistive wall allows rapid flux penetration, leading to an earlier crash.
	
	\section{Summary and discussion}\label{sec4}
	
	In this work, we have investigated the nonlinear disruption dynamics triggered by an RWM in a CFETR baseline steady-state equilibrium using the NIMROD code. The simulations reveal a deterministic disruption sequence in which the RWM evolves from a linear growth phase to a rapid macroscopic collapse. The onset of the thermal quench is precipitated by the global stochastization of magnetic field lines, which facilitates efficient parallel energy transport and leads to a catastrophic loss of core thermal energy. A characteristic transient current spike is observed near the end of the thermal quench; this feature is rigorously attributed to the conservation of poloidal magnetic flux during the rapid crash of the internal inductance. The subsequent current quench phase ensues as resistive dissipation increases sharply following the temperature collapse. The topological evolution demonstrates that confinement degradation is governed by the global unraveling of flux surfaces, consistent with an external-kink nature modulated by the resistive wall, rather than by a tearing-mode-like sequence driven by localized reconnection.
	
	Beyond the canonical quench sequence, a salient phenomenon of magnetic flux healing is observed during the deep current quench stage, where closed flux surfaces spontaneously re-form from the core outward. This recovery is accompanied by a vanishing pressure gradient and the flattening of the parallel current profile, indicating a self-organization toward a force-free equilibrium state consistent with Taylor relaxation. Sensitivity studies further show that the disruption timing is strongly modulated by nonlinear mode coupling, plasma resistivity, parallel thermal transport, and wall response. Including higher-$n$ toroidal harmonics accelerates the disruption onset via nonlinear mode coupling, whereas increased plasma resistivity and a more ideal wall delay it; stronger parallel heat transport shortens the thermal-collapse timescale once stochasticity develops.
	
	These results establish an MHD-level baseline for RWM-triggered disruptions in CFETR-relevant scenarios; however, quantitative prediction for reactor conditions necessitates physics extensions beyond the single-fluid framework. First, energetic particles (fusion-born alphas and beam ions) are expected to interact with low-frequency global modes through kinetic resonances, potentially modifying the stability thresholds and nonlinear saturation amplitudes relative to the purely thermal fluid description. Second, the post-disruption environment is critical for assessing machine safety, particularly regarding runaway electron generation. While the present model captures the macroscopic electromagnetic evolution, it does not account for the kinetic avalanche sources or the feedback of RE currents on the magnetic topology. Incorporating self-consistent energetic particle and runaway electron physics will be essential for establishing a predictive link between RWM-driven macroscopic dynamics and reactor-relevant hazard metrics in future studies.

	\ack
	We are grateful for the support from the NIMROD team. This work is supported by the National MCF Energy R\&D Program of China under Grant No.~2019YFE03050004, the U.S. Department of Energy Grant No.~DE-FG02-86ER53218, and the Hubei International Science and Technology Cooperation Project under Grant No.~2022EHB003. The computing work in this paper is supported by the Public Service Platform of High Performance Computing by Network and Computing Center of HUST.
	\section*{References}
	\bibliography{sample-1}
	\clearpage
	\clearpage
	\clearpage
	\clearpage
	\clearpage
	\clearpage
	\clearpage
	\clearpage
	\clearpage
	
	\clearpage
	\clearpage
	\clearpage
	\clearpage
\end{document}